\documentclass[conference]{IEEEtran}
\usepackage{cite}
\usepackage{amsmath,amssymb,amsfonts}
\usepackage{graphicx}
\usepackage{textcomp}
\usepackage{xcolor}
\usepackage{listings}
\usepackage{subcaption}
\usepackage{array}
\usepackage{multirow}
\usepackage[bookmarks=false]{hyperref}
\usepackage{algorithm}
\usepackage{algpseudocode}
\usepackage{enumitem}

\newcommand{\code}[1]{\texttt{#1}}

\def\tool{JCoffee}

\def\BibTeX{{\rm B\kern-.05em{\sc i\kern-.025em b}\kern-.08em
    T\kern-.1667em\lower.7ex\hbox{E}\kern-.125emX}}

\definecolor{codegreen}{rgb}{0,0.6,0}
\definecolor{codegray}{rgb}{0.5,0.5,0.5}
\definecolor{codepurple}{rgb}{0.58,0,0.82}
\definecolor{backcolour}{rgb}{0.95,0.95,0.92}
\definecolor{gray}{rgb}{0.4,0.4,0.4}
\definecolor{darkblue}{rgb}{0.0,0.0,0.6}
\definecolor{cyan}{rgb}{0.0,0.6,0.6}

\lstdefinelanguage{MyErrors}
{
  basicstyle=\ttfamily\footnotesize,
  sensitive=false,
  columns=fullflexible,
  showstringspaces=false,
  commentstyle=\color{gray}\upshape,
  morestring=[b]",
  morestring=[s]{>}{<},
  morecomment=[s]{<?}{?>},
  stringstyle=\color{black},
  numbersep=5pt,
  identifierstyle=\color{black},
  keywordstyle=\color{black},
  morekeywords={xmlns,version,type,symbol,location,error}
}

\lstdefinelanguage{Pseudocode}{
  keywords={typeof, new, true, false, catch, function, return, null, catch, switch, var, if, to, equals, in, while, do, for, else, case, break, write, compile, generate},
  ndkeywords={SUCCESS, FAILURE, class, export, boolean, throw, implements, import, this, jimple},
  ndkeywordstyle=\color{red},
  identifierstyle=\color{black},
  sensitive=false,
  backgroundcolor=\color{backcolour},   
  commentstyle=\color{codegreen},
  keywordstyle=\color{magenta},
  numberstyle=\tiny\color{codegray},
  stringstyle=\color{codepurple},
  basicstyle=\ttfamily\footnotesize,
  breakatwhitespace=false,
  breaklines=true,                 
  captionpos=b,                    
  keepspaces=true,                 
  numbers=left,                    
  numbersep=5pt,                  
  showspaces=false,                
  showstringspaces=false,
  showtabs=false,                  
  tabsize=2,
  comment=[l]{//},
  morecomment=[s]{/*}{*/},
  morestring=[b]',
  morestring=[b]"
}

\begin{document}
\newcommand{\STAB}[1]{\begin{tabular}{@{}c@{}}#1\end{tabular}}

\title{\tool: Using Compiler Feedback to Make Partial Code Snippets Compilable}

\author{
\IEEEauthorblockN{1\textsuperscript{st} Piyush Gupta}
\IEEEauthorblockA{\textit{Dept. of Computer Sc. and Engg.} \\
\textit{IIIT-Delhi}\\
New Delhi, India \\
piyush16066@iiitd.ac.in}
\and
\IEEEauthorblockN{2\textsuperscript{nd} Nikita Mehrotra}
\IEEEauthorblockA{\textit{Dept. of Computer Sc. and Engg.} \\
\textit{IIIT-Delhi}\\
New Delhi, India \\
nikitam@iiitd.ac.in }
\and
\IEEEauthorblockN{3\textsuperscript{rd} Rahul Purandare}
\IEEEauthorblockA{\textit{Dept. of Computer Sc. and Engg.} \\
\textit{IIIT-Delhi}\\
New Delhi, India \\
purandare@iiitd.ac.in}
}

\maketitle

\begin{abstract}
Static program analysis tools are often required to work with only a small part of a program’s source code, either due to the unavailability of the entire program or the lack of need to analyze the complete code. This makes it challenging to use static analysis tools that require a complete and typed intermediate representation (IR). We present \tool, a tool that leverages compiler feedback to convert partial Java programs into their compilable counterparts by simulating the presence of missing surrounding code. It works with any well-typed code snippet (class, function, or even an unenclosed group of statements) while making minimal changes to the input code fragment. A demo of the tool is available here: \url{https://youtu.be/O4h2g_n2Qls}
\end{abstract}
\begin{IEEEkeywords}
Partial programs, Compiler feedback, Static analysis, Type inference, Java
\end{IEEEkeywords}

\section{Introduction \& Motivation} \label{sec:Intro}

Static analysis is an indispensable tool to model a program's structure and analyze its behavior. Most static analysis tools work on an Intermediate Representation (IR) of the code, which is a trivial task for the compiler to build when the complete program is available at their disposal. However, when only a subset of the entire code is available, despite its syntactical correctness, static analysis tools refuse to operate on it due to ambiguities and references to undeclared constructs.

We define a partial program (PP) as a non-empty subset of an otherwise complete program (CP), which is unlikely to have any syntactic errors. A PP is incomplete because it may contain references to classes, methods, and variables whose declarations lie in CP but not in PP. Figure \ref{fig:partial_code_example} illustrates one such code snippet. Given only the code for a Java class \code{Bar}, we call it a partial program since it is ambiguous to determine the type of variable \code{lbl}, the parameter type of function \code{doB(\ldots)}, and the return type of function \code{doA()}.

The ability to analyze a partial program is advantageous in many scenarios where either the entire program is unavailable, or it is interesting to analyze only a recently-modified snippet of the entire program, such as:
\begin{itemize}
    \item encountering missing/deprecated dependencies during maintenance of an old project,
    \item analyzing code snippets from forum threads and documentation files before incorporating them in a codebase,
    \item independently checking bug fixes before automatic patching, and
    \item analyzing changes to code in web repositories to understand the evolution of source code.
\end{itemize}

\begin{figure}[tb]
    \centering
\begin{lstlisting}[language=Pseudocode, mathescape=true]
class Bar {
	void main() {
		Foo foo = new Foo();
		foo.lbl = "Hello world";
		foo.doB(foo.doA());
	}
}
\end{lstlisting}
    \caption{A Partial program (PP).}
    \label{fig:partial_code_example}
\end{figure}

In order to be able to analyze partial programs, it is essential to overcome the challenges, namely:

\begin{enumerate}[label=\roman*.]
	\item Resolving syntactical uncertainty that arises due to missing declarations. For example, \texttt{Foo.baz()} may denote both a call to a static function \texttt{baz()} of a class \texttt{Foo}; or a virtual method call to a member function \texttt{baz()} of an object \texttt{Foo} of some class, and

	\item Assigning correct data types for local variables, method return values, and class-level fields to enable IR generation.
\end{enumerate}

While complete and sound disambiguation of a partial program is an undecidable problem, it works in our favor that software engineering techniques can often trade some guarantees on correctness for increased precision \cite{ppa}. Some of the inferences we make during disambiguation may not be totally sound, but they provide little to no threat to the validity of the use case at hand.

The contributions of this paper are:
\begin{itemize}
	\item A novel approach for partial program analysis that infers missing code declarations based on the compiler's feedback during compilation in an iterative process. We focus on making little to no changes to the input snippet, but on adding other class/method definitions around it to complete the missing pieces.

	\item Our implementation - Java COmpiler For FEEdback (\tool{}) that realizes our approach.

	\item An evaluation of our tool on 9133 partial code snippets. \tool{} successfully disambiguated over 90\% of the inputs, with an overall accuracy of 93.3\%.

\end{itemize}

Our approach offers several benefits over existing methods. First, we do not need to propagate inferred information across errors since subsequent compile cycles automatically incorporate it. Second, we support a much broader use case as we do not need to scrape the web or maintain a repository of existing method signatures to match undeclared methods in input code. Lastly, as opposed to a simplified intermediate representation, we build the complete bytecode satisfying the strictest type guarantees of the compiler.

% The remainder of this paper is structured as follows:

\section{About \tool{}}  \label{sec:About}

\tool{} is based on the simple premise that Java compiler error messages are very verbose and, a lot of context needed to fix a given error is already present in its description.

Consider, for example, the PP in Figure \ref{fig:partial_code_example}. The errors generated during its compilation can be fixed by creating the class \code{Foo}, its member variable \code{lbl}, and functions \code{doA()} and \code{doB()} with the right signatures. Some of these errors and the corresponding fixed CP is shown in Figure \ref{fig:PP_errors_fixed}. The CP is now used to generate a class file, an intermediate representation (IR), a program dependence graph (PDG), or any other representation required for static analysis.

Interestingly, for each error, the compiler points out the exact location of the cause, the keywords it was expecting, and the keywords it found. Moreover, the errors are reported hierarchically, the fine-grained errors are detected only after the higher level inconsistencies have been resolved.
 
Thus, in a use-case such as ours, where the syntactic correctness of the code snippet is assured, it would not be wrong to make decisions based solely on the suggestions put forth by the compiler. 
% \rp{We need to clarify that we do not claim that the semantic correctness of a given PP does not necessarily refer to the semantics of the original PP if it exists. The semantic correctness is with respect to some CP that we eventually construct. If we do not clarify this it may set wrong and unreasonable expectations. We should make this clear upfront - perhaps here itself.}

Our only assumption apart from a well-typed input code is the availability of standard Java types, such as \texttt{java.lang.*}, or \texttt{java.io.*} during compile time. We view this as a reasonable assumption because they ship with standard compilers and are necessary to run Java programs.
The various errors fixed by \tool{} can be classified as:

\begin{itemize}
    \item Identifier errors
    
    \begin{itemize}
        \item \textit{cannot find symbol}: The commonest error, it is thrown whenever the compiler cannot find the declaration of an identifier (such as package, interface, class, method, constructor, or variable). \tool{} declares a new identifier with the corresponding missing signature.

        \item \textit{array required but \ldots{} found}: This is thrown when there is an attempt to index a variable that has not been declared as an array. \tool{} infers this and modifies the declared type of the variable.
    \end{itemize}
    
    \item Computation errors
    
    \begin{itemize}
        \item \textit{incorrect method, \ldots{} cannot be applied to \ldots{}}: This error results from an incompatibility between a method's call and its declaration. JCofee adapts the declaration to match the call signature.

        \item \textit{operator \ldots{} cannot be applied to \ldots{} / incompatible types / inconvertible types}: Some operators are only defined for specific types, and \tool{} uses these errors to infer the type of one of the participating operand whenever possible.

        \item \textit{invalid method declaration; return type required}: Every method must have a return type or void specified. \tool{} uses this fact to infer that the `method' here is in fact a constructor of the enclosing class.
    \end{itemize}
    
    \item Access to static entities
    
    \begin{itemize}
        \item \textit{non-static variable/method \ldots{} cannot be referenced from a static context}: The `static' modifier states that a variable/method is associated with a class, not individual objects. This helps \tool{} infer that the corresponding variable/method is to be labeled as `static'.
    \end{itemize}
    
    \item Miscellaneous errors
    
    \begin{itemize}
        \item \textit{for-each not applicable to expression type}: `for-each' is one of the many ways to iterate over an iterable in Java. This error indicates that the expression is an iterable, most likely an array. Repeated occurrences of this error in the same code fragment indicates the multidimensionality of the array.

        \item \textit{exception \ldots{} is never thrown in body of corresponding try statement}: This error occurs when the code snippet is missing the exception declaration, and \tool{} erroneously made a class by the same name on encountering a missing symbol error earlier. This error is then used to declare an exception instead. 
    \end{itemize}
\end{itemize}

\begin{figure}[t]
    \centering
    \begin{subfigure}[t]{0.45\textwidth}
        \centering
\begin{lstlisting}[language=MyErrors, mathescape=true]
error: cannot find symbol
  symbol:   class Foo
  location: class Bar

  symbol:   variable lbl
  location: variable foo of type Foo

  symbol:   method doA()
  location: variable foo of type Foo
\end{lstlisting}
        \caption{Simplified error log.}
    \end{subfigure}
    \begin{subfigure}[t]{0.45\textwidth}
        \centering
        \captionsetup{justification=centering}
\begin{lstlisting}[language=Pseudocode, mathescape=true]
class Bar {
	void main() {
		Foo foo = new Foo();
		foo.lbl = "Hello world";
		foo.doB(foo.doA());
	}
}

// Code added by JCoffee begins

class Foo {
	public String lbl;
	public UNKNOWN doA() {return null;}
	public void doB(UNKNOWN obj) {}
}

class UNKNOWN {
}
\end{lstlisting}
        \caption{Fixed complete program. \code{class UNKNOWN} instantiates all objects whose type cannot be resolved.}
    \end{subfigure}
    \caption{Errors and their fixes for a partial program.}
    \label{fig:PP_errors_fixed}
\end{figure}

\section{Architecture} \label{sec:Architecture}

At its core, \tool{} attempts to fix all the errors pointed out by the compiler in a deterministic manner, by simulating an environment of missing dependencies around the given code snippet. The tool consists of two modules, namely the \textbf{Error Fixing Module} (EFM) and the \textbf{Intermediate Representation Generator} (IRG). It also comprises the \textbf{Driver Engine} (DE) to integrate the two modules. Although the ideas presented are quite generic and may be implemented in any way suitable to the use case, we have implemented DE, EFM and IRG in python. We now discuss each %of the three components in detail.
component in detail.

\paragraph{Error Fixing Module (EFM)} Given a code snippet and a list of compiler-generated error messages, EFM fixes each error by adding class/method/variable declarations or inferring and updating the missing return type/declared type for methods/variables. Once all errors have been handled, EFM returns the modified code.

\begin{algorithm}[H]
    \caption{\tool{}}
    \label{alg:pseudocode}
    \begin{algorithmic}
        \Function{DE}{$code,maxSteps$}
            \State $code \gets preprocess(code)$
            \For{$ctr \gets 1$ to $maxSteps$}
                \State $nErr, errorMsgs \gets compile(code)$
                \If{$nErr = 0$}
                    \State \Call{IRG}{$code$}
                    \State \Return {$true$}
                \EndIf
                \State $code \gets$ \Call{EFM}{$code, errorMsgs$}
            \EndFor
            \State \Return{$false$}
        \EndFunction
        \\
        \Function{EFM}{$code,errorMsgs$}
            \State $modCode \gets code$
            \For{$err$ in $errorMsgs$}
                \State $modCode \gets fixError(code, err)$
            \EndFor
            \State \Return{$modCode$}
        \EndFunction
        \\
        \Function{IRG}{$code$}
            \State $outFile \gets code$
            \State $compile(outFile)$ \Comment{Used for static analysis}
        \EndFunction
    \end{algorithmic}
\end{algorithm}

\paragraph{Intermediate Representation Generator (IRG)} Once all the compiler-generated errors have been fixed, IRG generates bytecode or the \texttt{.class} files. The bytecode can be used as it is or further intermediate representations may be generated depending on the individual use case.

\paragraph{Driver Engine (DE)} The first step is to pre-process the input. If the input code snippet is not already a class (or a set of classes), it is encapsulated into a placeholder method and a class structure. DE then invokes \texttt{javac}, the Java compiler, on the input, to get a list of error messages which are sent to EFM. A successful compilation invokes IRG. Otherwise, it loops for a fixed number of iterations, with each EFM output becoming the input for the next invocation.

We outline the pseudocode for each of the components in Algorithm \ref{alg:pseudocode}. The simplistic and modular design of \tool\ makes it easy to extend it to support a wider range of errors, keeping it compatible with future Java versions.

\section{Evaluation} \label{sec:Evaluation}

We evaluated our implementation of \tool{} to answer the following research questions: \\
(RQ1): How effective is \tool{} in disambiguating code snippets? \\
(RQ2): What is the accuracy of \tool{} in preserving the intended semantics of code during disambiguation? \\
(RQ3): What is the time complexity and overhead of the iterative approach used in \tool{}?

\subsection{Dataset \& Environment}
We used 9133 partial code snippets from the BigCloneBench \cite{bcb} dataset released by Svajlenko et al. Ranging from a few statements (3 lines) to functions as large as 920 lines of code, the snippets contained code for frequently used functionalities from open source java projects. As almost all test samples referenced undeclared objects and functions, less than 1\% of the snippets were initially compilable.

The experiments were conducted on a server having OpenJDK Java 8, running Ubuntu 19.04 with 12 Intel Xeon E4 CPU cores and 31 GB of memory.

\subsection{RQ1: Effectiveness}
\tool{}'s primary goal is to completely resolve all compiler errors to generate bytecode for the given partial code. With some samples producing as many as 291 errors, we had a success rate of just over 90\% - \tool{} completely disambiguated 8220 samples out of the 9133 input snippets when allowed to run for up to 10 compiler iterations per input. Further, 94\% of the total samples were reduced to 2 or fewer compiler errors, which may then be fixed manually.

Upon further inspection of snippets that \tool{} failed to fix, we found the following recurring constructs:
\begin{itemize}
	\item inner classes
	\item Java generics
	\item classes/functions as parameters
	\item lambda expressions
\end{itemize}
We consider this to be a limitation of the current implementation, and plan to fix this in future versions.

\subsection{RQ2: Accuracy}

In some cases, the fix for a snippet does not completely capture the intended logic of the PP.

Consider, for example, the statement
\begin{center}
\code{dst.transfer(src, 0, src.size());}
\end{center}

Here, \code{src} and \code{dst} are objects of class \code{Channel}. While the PP contains their declaration, it does not contain the definition of class \code{Channel}. So, \tool{} sets the 3\textsuperscript{rd} parameter type of \code{transfer()} and the return type of \code{size()} as \code{UNKNOWN}. While it is trivial for a programmer to guess that \code{size()} %most 
likely returns an \code{int}, \tool{} does not yet use this extra %additional 
information available in variable \& function names.

To evaluate this impact, we randomly selected 30 test snippets, and manually fixed them, considering language constructs wherever possible. We compared this golden standard with the fixes generated by \tool{} for the same partial programs. Our results show that 93.3\% of the type inferences were identical, and in merely 6.7\% of the cases, a human evaluator could make more precise assumptions.

Hence, one extension of our approach can be to extract linguistic information from identifier names wherever possible and benefit from it.

\subsection{RQ3: Time complexity \& overhead}
The complexity of the algorithm is bounded by $n_s$, the number of statements in the source file, $n_e$, the average number of errors detected per pass, and $p$, the number of compile passes required to disambiguate the code completely. The time complexity can thus be expressed as $O(p*(n_s+n_e))$.

Out of the fixed files, 91.3\% required up to 5 compile iterations, while 99.5\% were fixed within seven iterations. On average, it took 1.64s to fix a code snippet. We view this as an acceptable cost for complete disambiguation.

\section{Related Work}  \label{sec:RelatedWork}

Several works exist that parse incomplete code to support static analyses. To the best of our knowledge, we know of no other research project aiming to generate compilable code for Java, or directly leveraging compiler feedback.
% However, they limit to the production of an IR, bytecode generation is not among their objectives.

PPA introduced by Dagenais and Hendren \cite{ppa} is the closest in terms of research objectives. It predicts unknown bindings for partial programs by performing partial type inference and uses heuristics to resolve ambiguities. However, it is compatible with only Java 1.4, and is now deprecated.

Melo et al. \cite{psychec} develop PsycheC, with a similar goal as ours, but for the C language. They build a lattice structure for various standard C types to bind variables to them during constraint generation for type inference.

Zhong and Wang \cite{grapa} propose GRAPA that locates previously released code archives, and extracts resolved types and method signatures to generate System Dependency Graphs \cite{grapa_12} (SDG) for partial programs. While suitable for a niche use case of publicly released applications, this approach fails for partial programs in general. 

PARSEWeb \cite{ppa_19}, PRIME \cite{grapa_25}, and SemDiff \cite{ppa_8} work similarly by recommending method calls with matching signatures from arbitrary frameworks or mining specifications for extracted API calls from the code snippet. Approaches such as \cite{grapa_28} only try to resolve missing dependencies given explicit import statements, something which is rarely available in partial programs.

% Other less related work includes approaches (\cite{grapa_9}, \cite{ppa_9}, \cite{ppa_11}) that analyze only a part of the intermediate representation, which is generated from the complete program with access to all type declarations.

Further, many Integrated Development Environments (IDE) such as \cite{eclipse_jdt}, \cite{eclipse_scrapbook} generate a typed abstract syntax tree for incomplete code snippets. However, they throw errors when they encounter constructs with missing declarations.

\section{Conclusions and Future Work}  \label{sec:Conclusion}

We implemented \tool{}, a tool that makes use of the detailed error information by the compiler to generate bytecode for partial programs by simulating the presence of undeclared constructs. Based on our evaluations on thousands of open source partial code snippets, \tool{} completely removes errors from over 90\% of input files, with the modifications being identical to the gold standard in 93.3\% of the cases.

In the future, we plan to support complex Java 8 mechanisms and  add capabilities to extract information from object names for a more precise type-inference. Also, since our approach is general, it can be adapted to other programming languages with descriptive errors, such as Rust and Elm.
% \rp{Should we also say that in the future, we would like to target partial programs that consist of incomplete or illegal syntactical units such as the ones that exist in Q\&A websites including StackOverflow?}\pg{My view is that if the constraints on valid syntax are also removed, then the problem becomes too vague to solve meaningfully. I believe that Q\&A sites like Stackoverflow already contain syntactically correct codes(atleast in the answers).} \rp{It does contain a lot of code that cannot be compiled. e.g. having ellipsis etc. However, we need not add this if we are not comfortable. We should have some goof future work though, which you have already included.}

Finally, the implementation of \tool{}, benchmarks, and outputs are available at \url{https://github.com/piyush69/JCoffee}.

% \pagebreak

\section*{Acknowledgment}
This work is partly supported by Infosys Center for Artificial Intelligence at IIIT-Delhi, Department of Science and Technology (DST) (India), Science and Engineering Research Board (SERB), the Confederation of Indian Industry (CII), and Nucleus Software Exports Ltd.

\bibliographystyle{IEEEtran}
\bibliography{list}
\end{document}